\documentclass[12pt]{article}
\usepackage{a4}
\usepackage{amsfonts}
\usepackage{amsmath}
\usepackage{latexsym}
\usepackage{amsfonts}
\usepackage{graphics}
\usepackage{graphicx}
\usepackage{epic}
\usepackage{eepic}
\usepackage{color}
\usepackage{bbm}

\newcommand{\cellsize}{10}
\newlength{\cellsz} 
\newsavebox{\cell}
\sbox{\cell}{\begin{picture}(\cellsize,\cellsize)
\put(0,0){\line(1,0){\cellsize}}
\put(0,0){\line(0,1){\cellsize}}
\put(\cellsize,0){\line(0,1){\cellsize}}
\put(0,\cellsize){\line(1,0){\cellsize}}
\end{picture}}
\newcommand\cellify[1]{\def\thearg{#1}\def\nothing{}%
\ifx\thearg\nothing
\vrule width0pt height\cellsz depth0pt\else
\hbox to 0pt{\usebox{\cell} \hss}\fi%
\vbox to \cellsz{
\vss
\hbox to \cellsz{\hss$#1$\hss}
\vss}}
\newcommand\tableau[1]{\vtop{\let\\\cr
\baselineskip -16000pt \lineskiplimit 16000pt \lineskip 0pt
\ialign{&\cellify{##}\cr#1\crcr}}}

\begin{document}


\renewcommand{\theequation}{\arabic{section}.\arabic{equation}}
\thispagestyle{empty}
\vspace*{-1,5cm}
\noindent \vskip3.3cm

\begin{center}
{\Large\bf On Higher Spin Symmetries in $AdS_{5}$}\\
\vspace*{1 cm}
{\large R. Manvelyan ${}^{\dag}$, K. Mkrtchyan${}^{\flat}$, R. Mkrtchyan ${}^{\dag}$
and S. Theisen ${}^{\ddag}$}
\vspace*{0.5 cm}

${}^{\dag}${\small\it Yerevan Physics Institute, Alikhanian Br. Str. 2, 0036 Yerevan, Armenia}

${}^{\flat}${\small\it Scuola Normale Superiore and INFN, Piazza dei Cavalieri 7, 56126 Pisa, Italy}

${}^{\ddag}${\small\it Max-Planck-Institut f\"ur Gravitationsphysik,
Albert-Einstein-Institut,\\ Am M\"uhlenberg 1, 14476 Golm, Germany}\\
\medskip
{\small\tt manvel@physik.uni-kl.de; karapet.mkrtchyan@sns.it; mrl55@list.ru; stefan.theisen@aei.mpg.de}
\end{center}\vspace{2cm}

\bigskip
\begin{center}
{\sc Abstract}
\end{center}
\quad
A special embedding of the $SU(4)$ algebra in $SU(10)$, including both spin two and
spin three symmetry generators, is constructed. A possible five dimensional action
for massless spin two and three fields with cubic interaction is constructed.
The connection with the previously investigated higher spin theories
in $AdS_{5}$ background is discussed.
Generalization to the more general case of symmetries, including spins $2,3,\dots s$, is shown.

\newpage

\section{Introduction}

\quad\, Higher Spin gauge theories have different structure in different space-time dimensions.
The first example of a consistent fully nonlinear HS theory in four dimensions
was given in
\cite{Vasiliev:1990en}. Less is known for higher dimensions. In dimensions
higher than four Higher Spin theories are getting more complicated in general, allowing fields
of mixed symmetry type. At the same time, for the restricted spectra of only symmetric
fields, Vasiliev equations are available  for any space-time dimension \cite{Vasiliev:2003ev}.
They are defined unambiguously
and describe totally symmetric bosonic fields of all spins.

Recent progress in three dimensional $AdS$ higher spin gravity resulted in
new relations between topological Chern-Simons theory, two-dimensional
conformal field theories with higher spin symmetry, and new three-dimensional
black hole solutions with higher spin charges (\cite{Campoleoni:2012hp}-\cite{Kraus:2011ds}
and references therein). It also points out again the importance of an $AdS$ background
for the construction of consistent nonlinear higher spin interactions with
a finite number of interacting higher spin gauge fields. These recent results are
based on the embedding of the gravitational gauge group into a larger group,
unifying higher spin gauge symmetry
with the $AdS$ group. In the three dimensional case it amounts to
embedding $SL(2)$ into $SL(3)$($SL(n)$) in the case of spin three
(up to spin $n$) gravity, and the corresponding field theory is described by a
three-dimensional Chern-Simons action with $SL(3)\times SL(3)$
($SL(n)\times SL(n)$) gauge group. The case of three dimensions is singled out by the
existence of a one-parameter family of Higher Spin algebras that underlie the
construction of Chern-Simons actions for the gauge fields
\cite{Blencowe:1988gj,Bergshoeff:1989ns,Vasiliev:1989re,Prokushkin:1999gc}
and Vasiliev equations, describing the interaction of Higher Spin gauge fields
with scalar matter \cite{Prokushkin:1998bq}.

The main goal of this paper is to generalize this approach
to five dimensions, and to construct possible interacting theories 
(actually with cubic interaction)
with finite number of higher spin fields in an $AdS_{5}$ background.
Moreover \emph{we show the existence of a sequence of Lie algebras,
the generators of which can be identified with the generators of
Higher Spin gauge symmetries for a finite number of symmetric fields in
$(A)dS_5$, in analogy to the three dimensional case.}\footnote{These algebras
should correspond to the representations of $su(2,2)$ (the latter can serve as 
defining representations for these algebras) found in 
\cite{Fernando:2009fq} and 
should be discrete cases of the one-parameter family of algebras of
\cite{Boulanger:2011se}.}

As a realization of this idea we construct in the next section a special embedding
of the spin two and spin three symmetry generators in frame formalism into a
unifying  $SU(10)$ Lie algebra, where the spin two generators correspond to the
$SU(4)$ subalgebra and the spin three generators to the remaining part
of $SU(10)$. In Section 3 we construct gauge fields
and curvatures. The latter include interactions and self-interactions of
the spin-2 and spin-3 fields through the structure constant of $SU(10)$ algebra.
In the fourth section we construct an action with cubic interaction following the 
prescription of \cite{Vasiliev:2001wa} and \cite{Vasilev:2011xf} and using our $SU(10)$ 
gauge transformation and curvatures as a
realization of the unified spin $2$ and $3$ gauge field theory.
Generalization to any spin is discussed in Section 5.

It would of course  be interesting to construct a fully nonlinear
interacting $SU(10)$ invariant action.
The first idea which comes to mind is a five-dimensional
Chern-Simons action for the $SU(10)$ gauge field. This idea is also based
on the fact that unitary groups have an invariant third rank
symmetric tensor which provides an invariant trace for the construction of
the Chern-Simons action in five dimensions. But it is well known 
\cite{Chamseddine:1989nu}\cite{Chamseddine:1990gk}
that this action, even in the pure gravity case ($SO(6)$ gauge group)
leads to Gauss-Bonnet (Lovelock) gravity with a special combination of terms quadratic
and linear in curvatures and  without a propagator for spin two fluctuations
in an $AdS_{5}$ background. Higher Spin Chern-Simons gravity in 5d was discussed
in \cite{Engquist:2007kz}, where the authors considered also the dynamics
of linearized spin $3$ gauge fields. A different Lagrangian
formulation for theories of spin 2 and higher in an $AdS$ background in the
frame formulation is
the so-called MacDowell-Mansouri-Stelle-West formulation
\cite{MacDowell:1977jt,Stelle:1979aj} used by
Vasiliev for a perturbative analysis of interactions 
\cite{Vasiliev:2001wa,Bekaert:2005vh,Vasilev:2011xf}.
In Appendix B we discuss
a generalization of the coset construction of
\cite{MacDowell:1977jt,Stelle:1979aj} and introduce a 
compensator field living on the coset $SU(10)/SO(10)$.
Unfortunately our result is negative: this theory does not have a correct free field limit.

\section{Unification of spin 2 and 3 symmetries on $AdS_{5}$}
\setcounter{equation}{0}
Gravitational theories in frame formalism can be formulated as gauge theories.
Since our construction draws some of its motivation from the three dimensional
case, we will briefly recall it. There pure gravity with a negative cosmological
constant can be written as a $SO(2,2)\simeq SL(2,{\mathbb R})\times SL(2,{\mathbb R})$
Chern-Simons theory. The generalization to higher spin is to replace $SL(2)$ by
a bigger group $G$ with a special embedding $SL(2,{\mathbb R})\hookrightarrow G$,
the simplest case being $G=SL(3,{\mathbb R})$ with the principal embedding, leading
to a unified description of a spin-three field coupled to gravity.

Five dimensional gravity in $AdS_{5}$ space is a gauge theory of $SO(2,4)$ (pure AdS)
or $SO(1,5)$ (Euclidian AdS). The corresponding f\"unfbein and spin connection
can be extracted from the gauge field, which is an algebra-valued one-form, by decomposition
of the adjoint representation of  $SO(2,4)$ or $SO(1,5)$
into the adjoint and vector representations of $SO(1,4)$. For simplicity and without loss of
generality we can replace these non-compact groups by their compact versions.
Namely we consider instead of the $AdS_5$ group the six dimensional rotation group $SO(6)$
and expand the gauge field with respect to the ``space-time rotation" group $SO(5)$,
just separating the sixth component as the vector representation and obtaining
correspondingly a f\"unfbein and a spin-connection:
\begin{eqnarray}\label{1.2}
  A^{AB}_{\mu}dx^{\mu}&=&A^{AB}= -A^{BA},\quad A,B,\dots=1,\dots, 6 ,\nonumber\\
  A^{AB} &=&\{A^{a6}, A^{ab}\}= \{e^{a}, \omega^{ab}\}, \quad a,b=1,\dots, 5 .
\end{eqnarray}
We can then impose constraints of vanishing torsion and express the spin connection
in terms of f\"unfbein and inverse f\"unfbein fields.

Then we propose the following extension to include spin 3 fields
(and higher).
The $SO(6)$ representation of the gravitational fields (\ref{1.2}) is via
the antisymmetric two cell Young tableau
\begin{equation}\label{1.3}
  {{A}}^{AB}\Rightarrow  Y^{SO(6)}_{{A}^{AB}}=\tableau{{\ }\\{\ }}\,\, ,
  \quad {\rm dim}(Y^{SO(6)}_{{  A}^{AB}})=15 ~.
\end{equation}
In terms of Young tableaux, the expansion (\ref{1.2}) is
\begin{equation}\label{1.4}
  \tableau{{\ }\\{\ }}_{~SO(6)}=\Big(~\tableau{{\ }}~+~\tableau{{\ }\\{\ }}~\Big)_{SO(5)} ~,
\end{equation}
or in terms of dimensions:
\begin{equation}\label{1.5}
\underline{\textbf{15}}_{SO(6)}=(\underline{\textbf{5}}+\underline{\textbf{10}})_{SO(5)} ~.
\end{equation}
From this point of view the spin 3 field corresponds to the $SO(6)$ window diagram
\cite{Vasiliev:2001wa}
\begin{equation}\label{1.6}
  {  A}^{AB,CD}\Rightarrow  Y^{SO(6)}_{{  A}^{AB,CD}}
  =\tableau{{\ }&{\ }\\{\ }&{\ }}\,\, , \quad\quad
  {\rm dim}\big(Y^{SO(6)}_{{  A}^{AB,CD}}\big)=84 ~.
\end{equation}
The conventions are such that $A$ is symmetric in each pair of indices. 
The corresponding $SO(5)$ expansion to a spin 3 tetrad and connections looks like
\begin{eqnarray}
  {  A}^{AB,CD} & & \quad e^{ab}\quad\quad\omega^{ab,c}~~~~ \omega^{ab,cd} \nonumber\\
  \tableau{{\ }&{\ }\\{\ }&{\ }}_{SO(6)} &=&\Big(~\tableau{{\ }&{\ }}
  ~+~ \tableau{{\ }&{\ }\\{\ }}~+~\tableau{{\ }&{\ }\\{\ }&{\ }}~\Big)_{SO(5)} ~,
  \label{1.7}\\
  \underline{\textbf{84}}_{SO(6)}&=&\big(~~\underline{\textbf{14}}~~\,
  +~~\underline{\textbf{35}}~+~~\underline{\textbf{35}}\big)_{SO(5)} ~.\nonumber
\end{eqnarray}
%
%
The $\omega^{ab,cd}$ are so-called extra fields (which are absent in $d=3$).

For the unification of the spin 2 and spin 3 degrees of freedom into one field,
we should first of all find a Lie group $G$ with dimension
\begin{equation}\label{1.10}
  \underline{\textbf{15}}_{SO(6)}+\underline{\textbf{84}}_{SO(6)}=\underline{\textbf{99}}_{\textbf{G}} ~.
\end{equation}
Taking into account that $SO(6)$ is equivalent\footnote{See
the appendix for details on the isomorphism $so(6)\simeq su(4)$ and other
relevant formulae.}
to $SU(4)$ we see that the natural choice for $G$ is $SU(10)$\footnote{For other
signatures of the initial space-time isometry algebra, we have, of course, different
real forms of $SL(10,\mathbb{C})$.}.
The $15$ generators of spin 2 gauge symmetry and $84$ generators of
spin 3 gauge symmetry can be combined into the 99 generators of $SU(10)$.

To proceed, we have to find an embedding of  $SU(4)$ into $SU(10)$ such that
the adjoint of the latter
decomposes w.r.t. the former as in (\ref{1.10}). That amounts to finding
a representation of $SU(4)$
of dimension $10$. Such representation of $SU(4)$ exists in the space of symmetric
second-rank tensors. We arrive at the following embedding procedure:\footnote{We do not
distinguish between the components of a tensor in the adjoint representation and the
generators of $SU(10)$.}
\begin{itemize}
\item Denote the $99$ generators of the $SU(10)$ algebra by
\begin{equation}\label{1.11}
  U^{I}_{J},\quad U^{I}_{I}=0, \quad I,J,\dots \in\{1,2,\dots, 10\}.
\end{equation}
\item We can present the $SU(10)$ vector indices $I,J,\dots$ as
symmetric pairs of vector indices of $SU(4)$
\begin{eqnarray}
  && I,J,\dots ~\to~(\alpha\beta),(\gamma\delta),\dots, \quad \alpha,\beta,\dots
  \in\{1,2,3,4\},\nonumber\\
\noalign{\vskip.2cm}
  &&U^{I}_{J}~~~\to~ U^{\alpha\beta}_{\gamma\delta} =U^{\beta\alpha}_{\gamma\delta}
  =U^{\alpha\beta}_{\gamma\delta},\quad U^{\alpha\beta}_{\alpha\beta}=0 .\label{1.111}
\end{eqnarray}
\item The $SU(4)\hookrightarrow SU(10)$ embedding can then be realized as the
decomposition into single and double traceless parts of $U^{\alpha\beta}_{\gamma\delta}$
\begin{eqnarray}
  U^{\alpha\beta}_{\gamma\delta}&=&W^{\alpha\beta}_{\gamma\delta}
  +\frac{1}{6}\delta^{(\alpha}_{(\gamma}L^{\beta)}_{\delta)},\label{1.12}\\
  L^{\beta}_{\delta}&=&U^{\alpha\beta}_{\alpha\delta},\nonumber\\
  W^{\alpha\beta}_{\alpha\delta}&=&L^{\beta}_{\beta}=0 ,\nonumber
\end{eqnarray}
where $L^{\beta}_{\delta}$ are the $15$ generators of $SU(4)$.
\end{itemize}
This shows that (\ref{1.12}) is a realization of the embedding:
\begin{equation}\label{1.16}
  \underline{\textbf{99}}_{SU(10)}=(\underline{\textbf{15}}+\underline{\textbf{84}})_{SO(6)}~.
\end{equation}
Using the explicit form of the $SU(10)$ generators, it is straightforward
to work out the commutation relations of $L$ and $W$. The result is
given in the appendix.

To summarize, we constructed a Lie algebra of spin
$3$ and spin $2$ transformations in $AdS_{5}$
using a special embedding $SO(6)\simeq SU(4)\hookrightarrow SU(10)$.
From \eqref{A.6}
one sees that the difference between $SU(10)$ and $SU(4)$ is precisely the tensor
representation of $SU(4)$ corresponding to the window tableau of $SO(6)$.

In the subsequent sections we attempt to construct  gauge field
theory with cubic interaction corresponding to the above unified algebra starting from Vasiliev's
free higher spin action in AdS background \cite{Vasiliev:2001wa}.

\section{Gauge fields and Curvatures}\setcounter{equation}{0}

\quad
In this section we apply the $SU(4)\hookrightarrow SU(10)$ embedding to
gauge fields and curvatures.
First of all we can equip a general one-form gauge field and zero-form gauge parameter
with $SU(10)$ indices expressed as symmetric pairs of $SU(4)$
indices
\begin{eqnarray}
  &&\textbf{A} = {  A}^{\alpha\beta}_{\gamma\delta}\,{U}^{\gamma\delta}_{\alpha\beta},\quad
  \epsilon = \epsilon^{\alpha\beta}_{\gamma\delta}\,{U}^{\gamma\delta}_{\alpha\beta}~,
  \label{2.1}\\
\noalign{\vskip.2cm}
  \delta \textbf{A}&=& D \epsilon \quad \Rightarrow \quad
  \delta{  A}^{\alpha\beta}_{\gamma\delta}= d\epsilon^{\alpha\beta}_{\gamma\delta}
  +{  A}^{\alpha\beta}_{\lambda\rho}\epsilon^{\lambda\rho}_{\gamma\delta}
  -{  A}^{\lambda\rho}_{\gamma\delta}\epsilon^{\alpha\beta}_{\lambda\rho}~. \nonumber
\end{eqnarray}
From now on we use for algebra valued objects a component formalism, i.e.
stripping off the generators. In this notation the $SU(10)$ Yang-Mills field strength is
\begin{equation}\label{2.3}
  {  F}^{\alpha\beta}_{\gamma\delta}=d{  A}^{\alpha\beta}_{\gamma\delta}
  +{  A}^{\alpha\beta}_{\lambda\rho}\wedge{  A}^{\lambda\rho}_{\gamma\delta},
  \quad {  F}^{\alpha\beta}_{\alpha\beta}=0 .
\end{equation}
Using the embedding (\ref{1.12}) we can extract from the $SU(10)$ gauge field
and field strength the spin $2$ and spin $3$ gauge fields and curvatures:
\begin{eqnarray}
  &&{  A}^{\alpha\beta}_{\gamma\delta}={  W}^{\alpha\beta}_{\gamma\delta}
        +\frac{1}{6}\delta^{(\alpha}_{(\gamma}\omega^{\beta)}_{\delta)}, \quad
        {  W}^{\alpha\beta}_{\alpha\delta}=\omega^{\beta}_{\beta}=0 ,\label{2.4}\\
  &&{  F}^{\alpha\beta}_{\gamma\delta}={  R}^{\alpha\beta}_{\gamma\delta}
        +\frac{1}{6}\delta^{(\alpha}_{(\gamma}r^{\beta)}_{\delta)}, \quad
        {  R}^{\alpha\beta}_{\alpha\delta}=r^{\beta}_{\beta}=0 .\nonumber
\end{eqnarray}
where
\begin{eqnarray}
  &&{  R}^{\alpha\beta}_{\gamma\delta} = D_{\omega}{  W}^{\alpha\beta}_{\gamma\delta}
  +{  W}^{\alpha\beta}_{\lambda\rho}\wedge {  W}^{\lambda\rho}_{\gamma\delta}
  -\frac{1}{6}\delta^{(\alpha}_{(\gamma}{  W}^{\beta)\sigma}_{|\lambda\rho|}\wedge
  {W}^{\lambda\rho}_{\delta)\sigma},  \nonumber\\
  && D_{\omega}{  W}^{\alpha\beta}_{\gamma\delta}=d{  W}^{\alpha\beta}_{\gamma\delta}
  +\frac{1}{3}\omega^{(\alpha}_{\lambda}\wedge{  W}^{\beta)\lambda}_{\gamma\delta}
  -\frac{1}{3}\omega^{\lambda}_{(\gamma}\wedge{  W}^{\alpha\beta}_{\delta)\lambda},\label{2.7}\\
  && r^{\alpha}_{\beta}=d\omega^{\alpha}_{\beta}
  +\frac{1}{3}\omega^{\alpha}_{\lambda}\wedge \omega^{\lambda}_{\beta}
  +{  W}^{\alpha\sigma}_{\lambda\rho}\wedge {  W}^{\lambda\rho}_{\beta\sigma}.\nonumber
\end{eqnarray}
Structure and couplings of fields in the curvatures reflect the structure
of the commutators \eqref{A.6}\footnote{After rescaling the spin two field
$\omega \rightarrow 3\, \omega$ the curvature takes the usual Riemann form.}.
Defining the $AdS_{5}$ background in standard $SU(4)$ covariant way as
\begin{eqnarray}
  && \omega^{\alpha}_{\mu}=\omega^{\alpha}_{0\mu} ,\\
  && r_0=D_{\omega_{0}}\omega_{0}=0 ,
\end{eqnarray}
where $D_{\omega_{0}}=d+\omega_{0}$ is the $AdS_{5}$ covariant exterior
derivative\footnote{
\begin{eqnarray}
&& D_{\omega_{0}}{  W}^{\alpha\beta}_{\gamma\delta}=d{  W}^{\alpha\beta}_{\gamma\delta}
  +\frac{1}{3}\omega^{(\alpha}_{0\lambda}\wedge{  W}^{\beta)\lambda}_{\gamma\delta}
  -\frac{1}{3}\omega^{\lambda}_{0(\gamma}\wedge{  W}^{\alpha\beta}_{\delta)\lambda},\nonumber\\
  &&D_{\omega_{0}}\omega^{\alpha}_{\beta}=d{  \omega}^{\alpha\beta}_{\gamma\delta}
  +\frac{1}{3}\omega^{\alpha}_{0\lambda}\wedge{  \omega}^{\lambda}_{\beta}
  -\frac{1}{3}\omega^{\lambda}_{0\beta}\wedge{  \omega}^{\alpha}_{\lambda} .\nonumber
\end{eqnarray}},
we can expand the gauge field in this background and extract from the $SU(10)$
field strength the spin $2$ and spin $3$ curvatures in both linear and
quadratic order in field fluctuations:
\begin{eqnarray}
  &&{  A}^{\alpha\beta}_{\gamma\delta}={  W}^{\alpha\beta}_{\gamma\delta}
        +\frac{1}{6}\delta^{(\alpha}_{(\gamma}(\omega_{0}+\omega)^{\beta)}_{\delta)},
        \\
  &&{  F}^{\alpha\beta}_{\gamma\delta}={  R}^{\alpha\beta}_{1\gamma\delta}+{  R}^{\alpha\beta}_{2\gamma\delta}
        +\frac{1}{6}\delta^{(\alpha}_{(\gamma}(r_{1}+r_{2})^{\beta)}_{\delta)} ,\nonumber
\end{eqnarray}
where
\begin{eqnarray}
&&{  R}^{\alpha\beta}_{1\gamma\delta}=D_{\omega_{0}}{  W}^{\alpha\beta}_{\gamma\delta},
\nonumber\\
&&{  R}^{\alpha\beta}_{2\gamma\delta} = \frac{1}{3}\omega^{(\alpha}_{\lambda}
\wedge{  W}^{\beta)\lambda}_{\gamma\delta}
-\frac{1}{3}\omega^{\lambda}_{(\gamma}\wedge{  W}^{\alpha\beta}_{\delta)\lambda}
+{  W}^{\alpha\beta}_{\lambda\rho}\wedge {  W}^{\lambda\rho}_{\gamma\delta}
-\frac{1}{6}\delta^{(\alpha}_{(\gamma}{  W}^{\beta)\sigma}_{|\lambda\rho|}\wedge
{W}^{\lambda\rho}_{\delta)\sigma},  \nonumber\\
&& r^{\alpha}_{1\beta}=D_{\omega_{0}}\omega^{\alpha}_{\beta},\nonumber\\
&&r^{\alpha}_{2\beta}=
\frac{1}{3}\omega^{\alpha}_{\lambda}\wedge \omega^{\lambda}_{\beta}
+{  W}^{\alpha\sigma}_{\lambda\rho}\wedge {  W}^{\lambda\rho}_{\beta\sigma}.
\label{2.8}
\end{eqnarray}
In the next section we construct a cubic interaction using these expansions.

\section{Spin 3 and 2 Cubic Interaction}
\setcounter{equation}{0}
To formulate correctly the free action,
we begin with a brief review of the Macdowell-Mansouri-Stelle-West action
principle for the case of ordinary spin two gravity in five dimensions.
The task can be formulated in the following way:
we have to write a topological action for a five dimensional gauge
theory with $SO(6)$ gauge group.
This means that we should construct a five-form enabling us to integrate over
a general five dimensional manifold $M_{5}$ in a metric independent way.
Introduce a field strength
\begin{equation}\label{3.1}
  {  F}^{AB}=d{  A}^{AB}+{  A}^{A}{}_{C}\wedge{  A}^{CB},\quad  A,B,\dots=1,2\dots 6 ~.
\end{equation}
The natural choice for the action is
\begin{equation}\label{3.2}
S_{SO(6)}\sim\int_{M_{5}}\epsilon_{ABCDEF}\,{B}^{AB}\wedge{  F}^{CD}\wedge{  F}^{EF} ~,
\end{equation}
where ${  B}^{AB}=-{  B}^{BA}$ is an $SO(6)$
algebra valued gauge covariant one-form constructed
from some compensator field. The compensator field should be introduced in a way
that does not lead to equations of motion purely quadratic in the field strength
\begin{equation}\label{3.3}
  \epsilon_{ABCDEF}{  F}^{CD}\wedge{  F}^{EF}=0,
\end{equation}
as happens in the Chern-Simons case and which leads to a vanishing propagator
in an $AdS$ background ${F}^{AB}={F}^{AB}_{AdS}=0$.
A possible solution is to take the compensator as an element of the coset  $G/H$ where $G$
in this case is $SO(6)$ and the stabilizer $H$ should be taken in a way to
keep ``Lorentz" covariance as the remaining symmetry after gauge fixing.
The natural choice in this case is $H=SO(5)$. This construction leads to a consistent
gravity action, which is equivalent to the Einstein-Hilbert action in the linearized limit.
In summary, we define the compensator field as an element of a five dimensional sphere
\begin{equation}\label{3.4}
S^{5}=SO(6)/SO(5) ~.
\end{equation}
The sphere can be realized, in a manifestly $SO(6)$ invariant way,
as a unit vector in ${\mathbb R}^6$:
\begin{equation}\label{3.5}
  {  V}^{A},  \quad {  V}^{A}{  V}_{A}=1 .
\end{equation}
The $SO(6)$ covariant one-form and the corresponding action can then be
constructed from (\ref{3.5}) uniquely:
\begin{eqnarray}
  {  B}^{AB} &=& {  V}^{[A}D{  V}^{B]},
  \quad D{  V}^{B}=d{  V}^{B}+{  A}^{B}_{\,\,\,\, C}V^{C} ,\label{3.6}\\
  S_{SO(6)} &\sim& \int_{M_{5}}\epsilon_{ABCDMN}\,
  {  V}^{A}D{  V}^{B}\wedge{  F}^{CD}\wedge{   F}^{MN} ~.\label{3.7}
\end{eqnarray}
A detailed analysis of the equations of motions and symmetries of this action can be
found in \cite{Vasiliev:2001wa}-\cite{LV}.
Here we only note that using local $SO(6)$ invariance of the theory, we can bring
the vector field ${  V}^{A}(x)$ to the constant unit vector in the sixth direction, and the
remaining $SO(5)$ invariance will still be sufficient for covariance in the language of
f\"unfbein and spin connection \eqref{1.2}. Another important aspect of this
construction is that the remaining $SO(5)$ invariance, combined with diffeomorphism 
invariance will still be sufficient for full AdS invariance of 
the theory \cite{Vasiliev:2001wa}.

The most important point of this short review for us is that
one can rewrite this action equivalently in $SU(4)$ form. This can be done
by direct transformation to chiral spinor indices
$\alpha,\beta, \dots\in\{1,2,3,4\}$ using standard identities for chiral
Dirac matrices in six dimensions\footnote{Further details are given in Appendix A.}
\begin{eqnarray}
&{V}^{\alpha\beta}= i (\Sigma^A)^{\alpha\beta}{  V}_A\quad\quad
\longleftrightarrow\quad {V}^A=\frac{i}{4}\Sigma^A_{\alpha\beta}{V}^{\alpha\beta}\,,\qquad
{V}^{\alpha\beta}=-{V}^{\beta\alpha} ,\nonumber\\
\noalign{\vskip.2cm}
&{F}^\beta_\alpha=(\Sigma_{AB})^\beta_\alpha{F}^{AB}\quad\longleftrightarrow\quad
{F}^{AB}=-{1\over2}(\Sigma^{AB})^\alpha_{\beta}{F}^{\beta}_{\alpha},
\qquad {F}^{\alpha}_{\alpha}=0 .\label{3.9}
\end{eqnarray}
The constraint on ${V}^{\alpha\beta}$ which follows from \eqref{3.5} is
\begin{equation}\label{3.10}
{V}^{\alpha\gamma}{V}_{\beta\gamma}=\delta^{\alpha}_{\beta},
\quad{  V}_{\alpha\beta}=\frac{1}{2}\epsilon_{\alpha\beta\gamma\delta}V^{\gamma\delta} .
\end{equation}
\setcounter{footnote}{0}
With the help of the identity \eqref{A.13} one obtains from \eqref{3.7}
\begin{equation}\label{3.13}
  S_{SU(4)} \sim i\int_{M_{5}}{V}^{\alpha\lambda}
  D{  V}_{\beta\lambda}\wedge {F}^{\beta}_{\rho}\wedge F^{\rho}_{\alpha} .
\end{equation}
So we recognize the $SU(4)$ covariant algebra-valued one-form\footnote{Another way of
transformating to the $SU(4)$ invariant action leading to the same result is
considered in Appendix B.}
\begin{eqnarray}
&&B^{\alpha}_{\beta}= i {V}^{\alpha\lambda}(D{V})_{\beta\lambda},
\quad B^{\alpha}_{\alpha}=0 ,\label{3.15}\\
&&(D{V})_{\beta\lambda}=d{V}_{\beta\lambda}
+{A}^{\rho}_{[\beta}{V}_{\lambda]\rho} .\nonumber
\end{eqnarray}
Linearization of this construction around an $AdS_{5}$ background
gives the free spin 2 action\footnote{Here and below, the overall normalization is
fixed for convenience.}
\begin{equation}\label{3.12}
 S_{SU(4)}^{s=2} = i\int_{M_{5}}{  V}^{\alpha\lambda}
  D_{\omega_{0}}{  V}_{\beta\lambda}\wedge {  r}^{\beta}_{1\rho}\wedge r^{\rho}_{1\alpha},
\end{equation}
which is the starting point for considering  free actions for higher spin fields in
$AdS_{5}$ space \cite{Vasiliev:2001wa}-\cite{LV}.
We now present the correct free action for spin three, which consist of two parts
\cite{Vasiliev:2001wa}
\begin{eqnarray}
S^{s=3}_{SO(6)} &\sim& \int_{M_{5}}\epsilon_{ABCDMN}\,
{V}^{A}D_{\omega_{0}}{V}^{B}\wedge\big(R_{1}^{CC_{1},DD_{1}}
\wedge R_1^{M}{}_{C_1}{}^{,N}{}_{D_1}\cr
\noalign{\vskip.2cm}
&&\qquad\qquad\qquad\qquad
+4\,{R_{1}}^{CC_{1},DD_{1}}\wedge R_{1}^{M}{}_{C_1}{}^{,ND_2}V_{D_{1}}V_{D_{2}}\big) ,
\end{eqnarray}
where the relative coefficient between the two terms is
fixed such that the equation of motion
for the unwanted ``extra" fields corresponding to the $SO(5)$ window
like Young tableau in (\ref{1.7}) trivializes.
Using results from Appendix A we can transform this action
to $SU(4)$ invariant form:
\begin{equation}
S^{s=3}_{SU(4)} =
i \int_{M^5}\,V^{\alpha \lambda}D_{\omega_{0}}V_{\mu\lambda}
\wedge\big(2\,R^{\,\mu\sigma}_{1\delta_1\delta_2}
\wedge R^{\,\delta_1\delta_2}_{1\alpha\sigma}
+R^{\,\mu\rho_{1}}_{1\sigma\delta_1}
\wedge R^{\,\sigma\rho_2}_{1\alpha\delta_{2}}V_{\rho_{1}\rho_{2}}
V^{\delta_{1}\delta_{2}}\big). \label{4.33}
\end{equation}
For the construction of the cubic interaction lagrangian using
our unifying spin 2 and 3 symmetry group $SU(10)$ we
start from the free spin 3 and spin 2 actions in $AdS_{5}$ background written 
in the $SU(4)$ form with an as yet undetermined relative coefficient $a$:
\begin{equation}\label{Sfree}
S^{\rm free}=a\, S_{SU(4)}^{s=2}+S_{SU(4)}^{s=3} .
\end{equation}
We then construct the cubic interaction following Noether's procedure and using
the $SU(10)$ transformations for curvatures (\ref{2.8}).
If we split the gauge parameter $\epsilon^{\alpha\beta}_{\mu\nu}$ into its spin tree
and two parts,
\begin{eqnarray}
\epsilon^{\alpha\beta}_{\mu\nu}&=&\eta^{\alpha\beta}_{\mu\nu}
+\frac{1}{6}\delta^{(\alpha}_{(\mu}\varepsilon^{\beta)}_{\nu)} ,\label{11}\\
\eta^{\alpha\beta}_{\alpha\nu} &=&\varepsilon^{\alpha}_{\alpha}=0 ,\nonumber
\end{eqnarray}
we derive the gauge transformation for the spin 3 and spin 2 curvatures\footnote{The
corresponding transformation for the gauge fields is:
\begin{eqnarray}
\delta W^{\alpha\beta}_{\mu\nu} &=& D_{\omega_{0}}\eta^{\alpha\beta}_{\mu\nu}
+[W,\eta]^{\alpha\beta}_{\mu\nu}
-\frac{1}{6}\delta^{(\alpha}_{(\mu}[W,\eta]^{\beta)\sigma}_{\nu)\sigma}
+\frac{1}{3}[W,\varepsilon]^{\alpha\beta}_{\mu\nu}
+\frac{1}{3}[\omega,\eta]^{\alpha\beta}_{\mu\nu} ,\\
\delta \omega^{\alpha}_{\mu} &=& D_{\omega_{0}}\varepsilon^{\alpha}_{\mu}+\frac{1}{3}
[\omega,\varepsilon]^{\alpha}_{\mu}+[W,\eta]^{\alpha\sigma}_{\mu\sigma}.
 \end{eqnarray}}:
\begin{eqnarray}\label{deltaR}
\delta R^{\alpha\beta}_{\mu\nu} &=&[R,\eta]^{\alpha\beta}_{\mu\nu}
-\frac{1}{6}\delta^{(\alpha}_{(\mu}[R,\eta]^{\beta)\sigma}_{\nu)\sigma}
+\frac{1}{3}[R,\varepsilon]^{\alpha\beta}_{\mu\nu}
+\frac{1}{3}[r,\eta]^{\alpha\beta}_{\mu\nu} ,\nonumber\\
\delta r^{\alpha}_{\mu} &=&\frac{1}{3}[r,\varepsilon]^{\alpha}_{\mu}
+[R,\eta]^{\alpha\sigma}_{\mu\sigma} ,
\end{eqnarray}
where
\begin{eqnarray}
&&[R,\eta]^{\alpha\beta}_{\mu\nu}
= R^{\alpha\beta}_{\lambda\rho}\eta^{\lambda\rho}_{\mu\nu}
-\eta^{\alpha\beta}_{\lambda\rho}R^{\lambda\rho}_{\mu\nu} ,\nonumber\\
&&[R,\varepsilon]^{\alpha\beta}_{\mu\nu}=
R^{\alpha\beta}_{\rho(\mu}\varepsilon^{\rho}_{\nu)}
-\varepsilon^{(\alpha}_{\rho}R^{\beta)\rho}_{\mu\nu} ,\\
&& [r,\eta]^{\alpha\beta}_{\mu\nu}=
r^{(\alpha}_{\rho}\eta^{\beta)\rho}_{\mu\nu}
-\eta^{\alpha\beta}_{\rho(\mu}r^{\rho}_{\nu)} ,\nonumber 
\end{eqnarray}
with
\begin{eqnarray}
[R,\varepsilon]^{\alpha\beta}_{\mu\beta} &=& [r,\eta]^{\alpha\beta}_{\mu\beta}=0 .
\end{eqnarray}
To perform Noether's procedure we split the gauge transformations
into zeroth and first order in gauge fields and, as typical for Yang-Mills type of
gauge fields, expand the transformations in first and second order in gauge fields
\begin{eqnarray}
&& \delta_{0} R^{\alpha\beta}_{1\mu\nu}=\delta_{0} r^{\alpha}_{1\mu}=0 ,\nonumber \\
&& \delta_{1} R^{\alpha\beta}_{1\mu\nu}+\delta_{0} R^{\alpha\beta}_{2\mu\nu}
={\Delta_{(R)}}^{\alpha\beta}_{\mu\nu}(R_{1},r_{1},\eta,\varepsilon) ,\nonumber\\
&&\delta_{1} r^{\alpha}_{1\mu}+\delta_{0} r^{\alpha}_{2\mu}
={\Delta_{(r)}}^{\alpha}_{\mu}(R_{1},r_{1},\eta,\varepsilon) ,\\
&&{\Delta_{(R)}}^{\alpha\beta}_{\mu\nu}=[R_{1},\eta]^{\alpha\beta}_{\mu\nu}
-\frac{1}{6}\delta^{(\alpha}_{(\mu}[R_{1},\eta]^{\beta)\sigma}_{\nu)\sigma}
+\frac{1}{3}[R_{1},\varepsilon]^{\alpha\beta}_{\mu\nu}
+\frac{1}{3}[r_{1},\eta]^{\alpha\beta}_{\mu\nu} ,\quad\quad\nonumber\\
&&{\Delta_{(r)}}^{\alpha}_{\mu}
=\frac{1}{3}[r_{1},\varepsilon]^{\alpha}_{\mu}
+[R_{1},\eta]^{\alpha\sigma}_{\mu\sigma} .\nonumber
\end{eqnarray}
We now use the prescription suggested in
\cite{Vasiliev:2001wa},\cite{Vasilev:2011xf} (see also \cite{Boulanger:2011se}
for further details and generalizations) and replace in the free action the
linearized curvatures by the full curvatures and extract a candidate
cubic action of the form:
\begin{eqnarray}\label{25}
&&S^{\rm cubic}= a i \int_{M^5}V^{\alpha \lambda}h_{\mu\lambda}
\wedge\Big[r^{\mu}_{2\delta}\wedge r^{\delta}_{1}{}_{\alpha}+ r^{\mu}_{1\delta}
\wedge r^{\delta}_{2\alpha}\Big]\nonumber\\
&&+\,i \int_{M^5}2 V^{\alpha \lambda}h_{\mu\lambda}
\wedge \Big[ R^{\mu\sigma}_{2\delta_1\delta_2}\wedge R^{\delta_1\delta_2}_{1\alpha\sigma}
+R^{\mu\sigma}_{1\delta_1\delta_2}\wedge R^{\delta_1\delta_2}_{2\alpha\sigma}\Big]\\
&&+\,i \int_{M^5} V^{\alpha \lambda}h_{\mu\lambda}
\wedge \Big[  R^{\mu\rho_1}_{2\sigma\delta_1}
\wedge R^{\sigma\rho_2}_{1\alpha\delta_{2}}+ R^{\mu\rho_1}_{1\sigma\delta_1}
\wedge R^{\sigma\rho_2}_{2\alpha\delta_{2}}\big]V_{\rho_{1}\rho_{2}}V^{\delta_{1}\delta_{2}}
\Big] ,\nonumber
\end{eqnarray}
where
\begin{eqnarray}
  h&=&D_{\omega_{0}}V,\quad D_{\omega_{0}}h=0\,.
\end{eqnarray}
This gives Noether's equation with nonzero right hand side
\begin{eqnarray}\label{26}
&&\delta_{1}S^{\rm free}+\delta_{0}S^{\rm cubic}
=a\, i\int_{M^5}V^{\alpha \lambda}h_{\mu\lambda}
\wedge\Big[\Delta_{(r)}^{\mu\delta}\wedge r^{\delta}_{1\alpha}+ r^{\mu}_{1\delta}
\wedge \Delta_{(r)\alpha}^{\delta}\Big]\nonumber\\
&&+\,i \int_{M^5}2\, V^{\alpha \lambda}h_{\mu\lambda}
\wedge \Big[ \Delta_{(R)\delta_1\delta_2}^{\mu\sigma}
\wedge R^{\delta_1\delta_2}_{1\alpha\sigma}
+R^{\mu\sigma}_{1\delta_1\delta_2}\wedge \Delta_{(R)\alpha\sigma}^{\delta_1\delta_2}\Big]\\
&&+\,i \int_{M^5} V^{\alpha \lambda}h_{\mu\lambda}
\wedge \Big[  \Delta_{(R)\sigma\delta_1}^{\mu\rho_1}
\wedge R^{\sigma\rho_2}_{1\alpha\delta_{2}}+ R^{\mu\rho_1}_{1\sigma\delta_1}
\wedge \Delta_{(R)\alpha\delta_{2}}^{\sigma\rho_2}\big]V_{\rho_{1}\rho_{2}} 
V^{\delta_{1}\delta_{2}}\Big]\, .\nonumber
\end{eqnarray}
It remains to prove that the right-hand side of Noether's equation
is zero on the free mass shell. This means that the r.h.s is zero on solutions
of the free equation of motion of the theory.
This requires a deformation of the initial Yang-Mills like gauge symmetry.
To show that r.h.s vanishes on the solutions of the free equations of motion
we use
the so-called First On-Shell Theorem \cite{Vasiliev:2001wa} which in our case
can be formulated in the following manner:
\begin{itemize}
\item All linearized `torsions' are zero on the free mass shell
\begin{eqnarray}
&& V^{\mu[\gamma}R^{\alpha]\beta}_{1\mu\nu}=V^{\mu[\gamma}r^{\alpha]}_{1\mu}=0 .\label{27}
\end{eqnarray}
\item The remaining curvatures can be expressed through the Weyl tensor zero
forms in the following way:
\begin{eqnarray}\label{28}
&& R^{\alpha\beta}_{1\mu\nu}=
H^{(2)}_{\lambda\rho}V^{\rho\gamma}C^{\lambda\alpha\beta}_{\gamma\mu\nu} ,\nonumber\\
&& r^{\alpha}_{1\mu}= H^{(2)}_{\lambda\rho}V^{\rho\gamma}c^{\lambda\alpha}_{\gamma\mu} ,\\
&& H^{(2)}_{\lambda\rho}=h_{\lambda\sigma}\wedge V^{\sigma\delta}h_{\delta\rho} .\nonumber
\end{eqnarray}
where the Weyl zero forms are completely symmetric and traceless
\begin{eqnarray}\label{31}
&&C^{\lambda\alpha\beta}_{\gamma\mu\nu}=C^{(\lambda\alpha\beta)}_{\gamma\mu\nu}
=C^{\lambda\alpha\beta}_{(\gamma\mu\nu)}, \quad c^{\lambda\alpha}_{\gamma\mu}
=c^{(\lambda\alpha)}_{\gamma\mu}
=c^{\lambda\alpha}_{(\gamma\mu)} ,\\
&&C^{\lambda\alpha\beta}_{\gamma\mu\beta}=c^{\lambda\alpha}_{\gamma\alpha}=0 .\nonumber
\end{eqnarray}
\item The Weyl tensors are $V$ transversal:
\begin{eqnarray}
&& V^{\rho[\delta}C^{\lambda]\alpha\beta}_{\rho\mu\nu}
=V^{\rho[\delta}c^{\lambda]\alpha}_{\rho\mu}=0 .\label{33}
\end{eqnarray}
\end{itemize}
The first simplification of the r.h.s. of \eqref{26} occurs by virtue of the
identity \eqref{A16}
and condition (\ref{27}). It allows us to remove the last line in (\ref{26}) while
changing the coefficient in the second line from 2 to 1.

A second simplification results from using the torsion free condition. It sets to zero
all terms in \eqref{26} which originate from the second term in $\Delta_{(R)}$
which effectively becomes
\begin{equation}\label{38}
  {\Delta_{(R)}}^{\alpha\beta}_{\mu\nu}=[R_{1},\epsilon]^{\alpha\beta}_{\mu\nu}
  +\frac{1}{3}[r_{1},\eta]^{\alpha\beta}_{\mu\nu}\,.
\end{equation}
Note that here the full parameter $\epsilon$ (cf. \ref{11}) appears.
The remaining terms can be written in the form
\begin{eqnarray}\label{39}
&&\delta_{1}S^{\rm free}+\delta_{0}S^{\rm cubic}\\
&&= a\, i \int_{M^5}V^{\alpha \nu}h_{\mu\nu}
\wedge\Big\{\frac{1}{3}[(r_{1}\wedge r_{1}),\varepsilon]^{\mu}_\alpha
+ [R_{1},\eta]^{\mu\sigma}_{\delta\sigma}\wedge r^{\delta}_{1\alpha}+r^{\mu}_{1\delta}
\wedge[R_{1},\eta]^{\delta\sigma}_{\alpha\sigma}\Big\}\nonumber\\
&&+\,i \int_{M^5} V^{\alpha \nu}h_{\mu\nu}
\wedge \Big\{[(R_{1}\wedge R_{1}),\epsilon]^{\mu\sigma}_{\alpha\sigma}
+ \frac{1}{3}[r_{1},\eta]^{\mu\sigma}_{\delta_1\delta_2}
\wedge R^{\delta_1\delta_2}_{1\alpha\sigma}+R^{\mu\sigma}_{1\delta_1\delta_2}
\wedge \frac{1}{3}[r_{1},\eta]^{\delta_1\delta_2}_{\alpha\sigma}\Big\} ,\nonumber
\end{eqnarray}
where
\begin{align}
& (R_{1}\wedge R_{1})^{\alpha\beta}_{\mu\nu}
=R^{\alpha\beta}_{1\rho_{1}\rho_{2}}\wedge R^{\rho_{1}\rho_{2}}_{1\mu\nu} ,\nonumber\\
\noalign{\vskip.2cm}
& (r_{1}\wedge r_{1})^{\alpha}_{\mu}=r^{\alpha}_{1\rho}\wedge r^{\rho}_{1\mu} .
\end{align}
Then inserting (\ref{28}) in (\ref{39}) we obtain
\begin{align}\label{42}
&\delta_{1}S^{\rm free}+\delta_{0}S^{\rm cubic}\nonumber\\
\noalign{\vskip.2cm}
&= i \int_{M^5} h_{\mu\nu}\wedge H^{(2)}_{\lambda\rho}\wedge H^{(2)}_{\phi\chi}
V^{\alpha\nu}V^{\rho\gamma}V^{\chi\tau}
\Big\{a \Big(\frac{1}{3}[(c^{\lambda}_{\gamma}c^{\phi}_{\tau}),\varepsilon]^{\mu}_\alpha
+ [C^{\lambda}_{\gamma},\eta]^{\mu\sigma}_{\delta\sigma}c^{\phi\delta}_{\tau\alpha}
+ c^{\phi\mu}_{\tau\delta} [C^{\lambda}_{\gamma},\eta]^{\delta\sigma}_{\alpha\sigma}\Big)
\nonumber\\
\noalign{\vskip.2cm}
&\qquad\qquad+[(C^{\lambda}_{\gamma}C^{\phi}_{\tau}),\epsilon]^{\mu\sigma}_{\alpha\sigma}
+\frac{1}{3}[c^{\lambda}_{\gamma},\eta]^{\mu\sigma}_{\delta_1\delta_2}
C^{\phi\delta_1\delta_2}_{\tau\alpha\sigma}
+\frac{1}{3}C^{\phi\mu\sigma}_{\tau\delta_1\delta_2}
[c^{\lambda}_{\gamma},\eta]^{\delta_1\delta_2}_{\alpha\sigma}\Big\}\, .
\end{align}
We now note the crucial identity
\begin{equation}
h_{\mu\nu}\wedge H^{(2)}_{\lambda\rho}\wedge H^{(2)}_{\phi\chi}
=\frac{1}{60}H^{(5)}\Big[V_{\lambda[\mu}V_{\nu](\phi}V_{\chi)\rho}
+V_{\rho[\mu}V_{\nu](\phi}V_{\chi)\lambda}+V_{\mu\nu}V_{\lambda(\phi}V_{\chi)\rho}\Big]\, ,
\end{equation}
where
\begin{equation}
H^{(5)}=V^{\chi\mu}h_{\mu\nu}\wedge V^{\nu\lambda} h_{\lambda\sigma}
\wedge V^{\sigma\delta}h_{\delta\rho}\wedge V^{\rho\phi}h_{\phi\delta}
\wedge V^{\delta\gamma}h_{\gamma\chi} 
\end{equation}
is a volume form.
Using this identity and the properties of the Weyl tensors, eqs. \eqref{31} and \eqref{33},
the variation of the action simplifies considerably:
\begin{align}
&\delta_{1}S^{\rm free}+\delta_{0}S^{\rm cubic}\nonumber\\
&= -{i\over 30}\Big(2a-{4\over3}\Big) \int_{M^5}H^{(5)} \Big(C^{\phi\mu\delta}_{\tau\theta_{1}\theta_{2}}
\eta^{\theta_{1}\theta_{2}}_{\delta\theta}c^{\tau\theta}_{\phi\mu}
- c^{\tau\theta}_{\phi\mu}\eta^{\mu\delta}_{\theta_{1}\theta_{2}}
C^{\phi\theta_{1}\theta_{2}}_{\tau\theta\delta}\Big)\, .
\label{50}
\end{align}
So we see that full cancelation occurs if we fix the coefficient
$a=\frac{2}{3}$.
We have thus shown that the invariant action with cubic interaction is
\begin{eqnarray}
&&S^{\rm free+cubic}= \frac{2}{3} i \int_{M^5}V^{\alpha \lambda}h_{\mu\lambda}
\wedge\Big[r^{\mu}_{1\delta}\wedge r^{\delta}_{1\alpha}
+r^{\mu}_{2\delta}\wedge r^{\delta}_{1\alpha}
+ r^{\mu}_{1\delta}\wedge r^{\delta}_{2\alpha}\Big]\nonumber\\
\noalign{\vskip.2cm}
&&+i \int_{M^5}2 V^{\alpha \lambda}h_{\mu\lambda}
\wedge \Big[R^{\mu\sigma}_{1\delta_1\delta_2}\wedge R^{\delta_1\delta_2}_{1\alpha\sigma}
+R^{\mu\sigma}_{2\delta_1\delta_2}\wedge R^{\delta_1\delta_2}_{1\alpha\sigma}
+R^{\mu\sigma}_{1\delta_1\delta_2}\wedge R^{\delta_1\delta_2}_{2\alpha\sigma}\Big]\\
\noalign{\vskip.2cm}
&&+i \int_{M^5} V^{\alpha \lambda}h_{\mu\lambda}
\wedge \Big[ R^{\mu\rho_1}_{1\sigma\delta_1}
\wedge R^{\sigma\rho_2}_{1\alpha\delta_{2}}+ R^{\mu\rho_1}_{2\sigma\delta_1}
\wedge R^{\sigma\rho_2}_{1\alpha\delta_{2}}+ R^{\mu\rho_1}_{1\sigma\delta_1}
\wedge R^{\sigma\rho_2}_{2\alpha\delta_{2}}\big]V_{\rho_{1}\rho_{2}}
V^{\delta_{1}\delta_{2}}\Big] .\nonumber
\end{eqnarray}
This action can be extracted as an expansion up to cubic order of the following
expression written in the form which includes only the $SU(10)$ field strength
$F^{\alpha\beta}_{\mu\nu}$:
\begin{eqnarray}
&& S^{\rm free+cubic}=i \int_{M^5}\Big\{ \frac{1}{3}V^{\alpha \lambda}
h_{\mu\lambda}\wedge F^{\mu\sigma}_{\rho\sigma}\wedge
F^{\rho\delta}_{\alpha\delta}\\
\noalign{\vskip.2cm}
&&\qquad\qquad + 2V^{\alpha \beta}h_{\mu\beta}F^{\mu\sigma}_{\lambda\rho}
\wedge F^{\lambda\rho}_{\alpha\sigma}+ V^{\alpha \lambda}h_{\mu\lambda}
\wedge F^{\mu\delta_{1}}_{\beta\rho_{1}}
\wedge F^{\sigma\delta_{2}}_{\alpha\rho_{2}}V^{\rho_{1}\rho_{2}}V_{\delta_{1}\delta_{2}}
\nonumber\\
\noalign{\vskip.2cm}
&&\qquad\qquad-\frac{4}{3}V^{\alpha \lambda}h_{\mu\lambda}F^{\mu\sigma}_{\alpha\rho}
\wedge F^{\rho\delta}_{\sigma\delta}
-\frac{2}{3}V^{\alpha \lambda}h_{\mu\lambda}F^{\mu\delta_{1}}_{\alpha\rho_{1}}
\wedge F^{\delta_{2}\sigma}_{\rho_{2}\sigma}V^{\rho_{1}\rho_{2}}
V_{\delta_{1}\delta_{2}}\Big\}\nonumber .\label{52}
\end{eqnarray}
With the help of identity (\ref{A.16}) we can rewrite (\ref{52}) as
\begin{eqnarray}
&& S^{\rm free+cubic}=
\frac{1}{3}i \int_{M^5}[V^{\alpha \lambda}h_{\mu\lambda}\delta^{\beta}_{\sigma}
-V^{\beta\alpha}h_{\mu\sigma}+h^{\beta\alpha}V_{\mu\sigma}]\\
&&\qquad\qquad\qquad\times\, (2\delta^{\rho_{1}}_{\delta_{2}}\delta^{\rho_{2}}_{\delta_{1}}
+V^{\rho_{1}\rho_{2}}V_{\delta_{1}\delta_{2}}
+\frac{1}{3}\delta^{\rho_{1}}_{\delta_{1}}\delta^{\rho_{2}}_{\delta_{2}})
\wedge F^{\mu\delta_{1}}_{\beta\rho_{1}}\wedge F^{\sigma\delta_{2}}_{\alpha\rho_{2}}.
\nonumber\label{3.59}
\end{eqnarray}
Analyzing this expression we find that the first bracket removes from the product of
two $SU(10)$ field strengths the quadratic term which mixes the spin two and spin three
fields. In the free limit this leads to the correct diagonal action
(\ref{Sfree}). On the other hand the operator
\begin{equation}\label{3.60}
2\delta^{\rho_{1}}_{\delta_{2}}\delta^{\rho_{2}}_{\delta_{1}}
+V^{\rho_{1}\rho_{2}}V_{\delta_{1}\delta_{2}} ,
\end{equation}
in the second bracket
controls the trivialization of the ``extra field" equation of motion for the
spin 3 part and the coefficient
$\frac{1}{3}$ in front of last term is fixed by the condition that a
deformation of the $SU(10)$ gauge invariance which leads to this cubic interaction
exists. This nontrivial deformation makes the generalization of this
procedure to quartic or the full nonlinear action illusive and as a
result the compact expression (\ref{3.59}) is correct only up
to cubic order.

One might consider avoiding the deformation and to
generalize the nonlinear spin 2 ($SU(4)$)
action (\ref{3.13}) to the spin 3 ($SU(10)$) case by introducing
a $SU(10)$ covariant compensator. But this does not provide the correct
free limit without mixed terms and the triviality of ``extra" field
free equations at the same time. This is demonstrated in Appendix B.

\section{Outlook}\setcounter{equation}{0}

One obvious generalization can be envisioned: including spins higher than three. This
generalization is straightforward as far as the identification of $G$ and the
embedding $SO(6)\hookrightarrow G$ are concerned. Consider
e.g. spin 2, spin 3 and spin 4. The fields and their $SO(5)$ representations are
\begin{align}
e^a~\tableau{{}}~\underline{\textbf{5}}~~
&& e^{ab}~\tableau{{}&{}}~\underline{\textbf{14}}
&& e^{abc}~\tableau{{}&{}&{}}~\underline{\textbf{30}}~~\nonumber\\
\noalign{\vskip.2cm}
\omega^{ab}~\tableau{{}\\{}}~\underline{\textbf{10}}
&& \omega^{ab,c}~\tableau{{}&{}\\{}&}~\underline{\textbf{35}}
&& \omega^{abc,d}~\tableau{{}&{}&{}\\{}& &}~\underline{\textbf{81}}~~\\
\noalign{\vskip.2cm}
&& \omega^{ab,cd}~\tableau{{}&{}\\{}&{}}~\underline{\textbf{35}}
&& \omega^{abc,de}~\tableau{{}&{}&{}\\{}&{}&}~\underline{\textbf{105}} \nonumber\\
\noalign{\vskip.2cm}
&& && \omega^{abc,def}~\tableau{{}&{}&{}\\{}&{}&{}}~\underline{\textbf{84}}~~\nonumber\\
\nonumber
\end{align}
The fields in each column combine into representations of $SO(6)$ whose
Young tableau coincides with the last one in each column.
The total of 399 fields nicely
combine into the adjoint representation of $SU(20)$.
The pattern repeats if we add higher spins such that for spin $2,\dots,s$ we find
$SU\big({s+2\choose3}\big)$.
All of the fields, that correspond to spins from 2 to $s$ now combine into one
$SU\big({s+2\choose3}\big)$-valued one-form master field. We can introduce
$s-1$ symmetrized $su(4)$ indices for each of the $SU\big({s+2\choose3}\big)$ indices
(the number of components matches exactly). The trace decomposition of the master
one-form field gives all the fields, corresponding to different spins.

We expect that this result hints on the existence of one parameter family of algebras
for symmetric Higher Spin fields in five dimensions, in full analogy with the three
dimensional case. For the critical values of the parameter, this algebra should acquire
infinite-dimensional ideals, with the remaining generators forming finite dimensional
subalgebras $SU\big({s+2\choose3}\big)$. This sequence of algebras should include
the known infinite dimensional Higher Spin algebras, discussed in
\cite{FradkinL,Vasiliev:2001wa,Sezgin:2001zs,Vasiliev:2003ev,Engquist:2007kz}.
In order to check this idea, one has to implement the more general construction of
Higher Spin algebra, along the lines of
\cite{Eastwood:2002su,Vasiliev:2004cm,Iazeolla:2008ix,Iazeolla:2008bp}.
In fact, a one parameter family of Higher Spin algebras is known to exist in any
dimension \cite{Boulanger:2011se}
(see also \cite{Boulanger:2013zza}).
This family of algebras includes mixed symmetry fields in higher dimensions,
while in five dimensions it does not.
It is also known \cite{Fernando:2009fq} that there is a family of
unitary representations of the $AdS_5$ algebra $su(2,2)$ that should serve as
defining representations for these algebras.

While we have demonstrated the correct cubic action for spin two and spin three in an AdS
background, we encounter standard problems when considering the fully interacting
theory, even in the case of our 
higher spin algebra with only finite number of spins (see Appendix
B for an alternative attempt).
Therefore, the question of existence of an action with nonlinear interactions of 
a finite number of dynamical Higher Spin fields remains open.

\section*{Acknowledgments}
\quad
This work is partially supported by Volkswagen Foundation.
R. Manvelyan and R. Mkrtchyan were partially supported by the grant of the
Science Committee of the Ministry of Science and Education of the Republic of Armenia
under contract 11-1c037 and 13-1C232. The work of R. Manvelyan was also partially
supported by grant ANSEF 2012. The work of K.M. was supported in part by
Scuola Normale Superiore, by INFN (I.S. TV12), by the MIUR-PRIN contract
2009-KHZKRX and by the ERC Advanced Investigator Grants no. 226455 Supersymmetry,
Quantum Gravity and Gauge Fields (SUPERFIELDS). K.M. would like to thank
Euihun Joung and Arthur Lipstein for useful comments. The authors are grateful to
S. Fredenhagen for helpful discussions and to N. Boulanger and E. Skvortsov
for comments.
R. Manvelyan and K. Mkrtchyan thank the
Galileo Galilei Institute for Theoretical Physics for the hospitality and
the INFN for partial support during the completion of this work.
We would also like to thanks the anonymous referee for his remarks and suggestions
which led to the revision of our original manuscript.

\section*{Appendix A: Useful Relations}
\setcounter{equation}{0}
\renewcommand{\theequation}{A.\arabic{equation}}

In this appendix we give some of the details about the Lie-algebras which were
used in the main body of the paper.

The generators of $SU(n)$ in the fundamental representation can be chosen
as a basis of real traceless matrices as follows:
\begin{equation}\label{A.1}
  (U^I_J)^i_j=\delta^{Ii}\delta_{Jj}-{1\over n}\delta^I_J\delta^i_j ,
\end{equation}
where the range of all indices is $1,\dots,n$. These generators satisfy
\begin{equation}
   [U^I_J,U^K_L]=\delta^K_J\, U^I_L-\delta^I_L\, U^K_J .\label{A.2}
\end{equation}
Using the explicit representation \eqref{A.1}, one easily works out the rank three
$d$-symbol of $SU(n)$:
\begin{align}\label{A.3}
d^{IKM}_{JLN}&={1\over 2}{\rm tr}(U^I_J\{U^K_L,U^M_N\})\\
&={1\over2}\Big(\delta^I_N\delta^M_L\delta^K_J
+\delta^I_L\delta^M_J\delta^K_N
-{2\over n}\delta^I_N\delta^K_L\delta^M_J
-{2\over n}\delta^M_L\delta^K_N\delta^I_J
-{2\over n}\delta^I_L\delta^K_J\delta^M_N
+{4\over n^2}\delta^I_J\delta^K_L\delta^M_N\Big) .\nonumber
\end{align}

Considering the special embedding $SU(4)\hookrightarrow SU(10)$, we represent the
$SU(10)$ indices $I,J,\dots$ by a symmetriced pair of $SU(4)$ indices, i.e.
$I=(\alpha\beta)$, etc. with $\alpha,\beta,\dots=1,\dots,4$ and rewrite
\eqref{A.2} as
\begin{equation}\label{A.4}
  [U^{\alpha\beta}_{\gamma\delta},U^{\mu\nu}_{\rho\sigma}]
  =\delta^{\mu\nu}_{\gamma\delta}\,U^{\alpha\beta}_{\rho\sigma}
  -\delta^{\alpha\beta}_{\rho\sigma}\,U^{\mu\nu}_{\gamma\delta}\,,\qquad
  \delta^{\alpha\beta}_{\gamma\delta}
  =\delta^\alpha_\gamma\delta^\beta_\delta+\delta^\alpha_\delta\delta^\beta_\gamma .
\end{equation}
Given the decomposition\footnote{Our conventions are
$\delta^{(\alpha}_{(\gamma}L^{\beta)}_{\delta)}
=\delta^\alpha_\gamma L^\beta_\delta+\delta^\beta_\gamma L^\alpha_\delta
+\delta^\alpha_\delta L^\beta_\gamma+\delta^\beta_\delta L^\alpha_\gamma$.}
\begin{equation}
  U^I_J=U^{\alpha\beta}_{\gamma\delta}=W^{\alpha\beta}_{\gamma\delta}
  +{1\over 6}\delta^{(\alpha}_{(\gamma}L^{\beta)}_{\delta)}\,,\qquad
  W^{\alpha\beta}_{\alpha\gamma}=L^\alpha_\alpha=0 ,
\end{equation}
and the algebra \eqref{A.4}, it is straightforward to derive
\begin{align}\label{A.6}
  &[L^\alpha_\beta,L^\gamma_\delta]=\delta^\gamma_\beta L^\alpha_\delta
  -\delta^\alpha_\delta L^\gamma_\beta ,\nonumber\\
\noalign{\vskip.2cm}
  &[L^\alpha_\beta,W^{\mu\nu}_{\rho\sigma}]
  =\delta^\alpha_{(\rho}W^{\mu\nu}_{\sigma)\beta}
  -\delta^{(\mu}_\beta W^{\nu)\alpha}_{\rho\sigma} ,\\
\noalign{\vskip.2cm}
  &[W^{\alpha\beta}_{\gamma\delta},W^{\mu\nu}_{\rho\sigma}]
  =\delta^{\mu\nu}_{\gamma\delta}W^{\alpha\beta}_{\rho\sigma}
  -\delta^{\alpha\beta}_{\rho\sigma}W^{\mu\nu}_{\gamma\delta}\nonumber\\
  &\qquad+{1\over6}\Big(\delta^{\alpha\beta}_{\langle\gamma(\rho}W^{\mu\nu}_{\sigma)\delta\rangle}
  -\delta^{\mu\nu}_{\langle\gamma(\rho}W^{\alpha\beta}_{\sigma)\delta\rangle}
  -\delta^{\langle\alpha(\mu}_{\gamma\delta}W^{\nu)\beta\rangle}_{\rho\sigma}
  +\delta^{\langle\alpha(\mu}_{\rho\sigma}W^{\nu)\beta\rangle}_{\gamma\delta}\Big)\nonumber\\
  &\qquad\quad
  +{1\over 6}\Big(\delta^{\mu\nu}_{\gamma\delta}\delta^{(\alpha}_{(\rho}L^{\beta)}_{\sigma)}
  -\delta^{\alpha\beta}_{\rho\sigma}\delta^{(\mu}_{(\gamma}L^{\nu)}_{\delta)}\Big)\nonumber\\
  &\qquad\qquad+{1\over 72}\Big(
  \delta^{\alpha\beta}_{\langle\gamma(\rho}\delta^{(\mu}_{\sigma)}L^{\nu)}_{\delta\rangle}
  -\delta^{\mu\nu}_{\langle\rho(\gamma}\delta^{(\alpha}_{\delta)}L^{\beta)}_{\sigma\rangle}
  -\delta^{\langle\alpha(\mu}_{\gamma\delta}\delta^{\nu)}_{(\rho}L^{\beta\rangle}_{\sigma)}
  +\delta^{\langle\mu(\alpha}_{\rho\sigma}\delta^{\beta)}_{(\gamma}L^{\nu\rangle}_{\delta)}\Big) ,
  \nonumber
\end{align}
where $\langle\alpha(\beta\gamma)\delta\rangle$ denotes symmetrization in $(\alpha,\delta)$
and in $(\beta,\gamma)$ and
$\delta^{\alpha\beta}_{\gamma\delta}=\delta^\alpha_\gamma\delta^\beta_\delta+
\delta^\alpha_\delta\delta^\beta_\gamma$.

The isomorphism between the vector respresentation of $SO(6)$ and the antisymmetric
second rank tensor representation of $SU(4)$ is made explicit with the help of
the chiral Dirac matrices, some of whose properties
are\footnote{The indices $\dot\alpha$ referring to the other chirality are not
needed here. By raising and lowering them with the charge conjugation matrix we can
always convert them to un-dotted indices.}
\begin{align}
& \Sigma^A_{\alpha\beta}=-\Sigma^A_{\beta\alpha} ,\nonumber\\
&(\Sigma^A)^{\alpha\beta}={1\over 2}\epsilon^{\alpha\beta\gamma\delta}\Sigma^A_{\gamma\delta} ,\\
&(\Sigma^A)^{\alpha\gamma}\Sigma^B_{\gamma\beta}+
(\Sigma^B)^{\alpha\gamma}\Sigma^A_{\gamma\beta}=2\delta^{AB}\delta^{\alpha}_\beta .\nonumber
\end{align}
A convenient basis for the $\Sigma^A_{\alpha\beta}$ is
$\Sigma^1=i\sigma_3\otimes\sigma_1,\,
\Sigma^2=\mathbbm{1}\otimes\sigma_2,\,
\Sigma^3=i\sigma_2\otimes\mathbbm{1},\,
\Sigma^4=\sigma_2\otimes\sigma_3,\,
\Sigma^5=i\sigma_1\otimes\sigma_2,\,
\Sigma^6=\sigma_2\otimes\sigma_1$ where $\sigma_i$
are the three Pauli matrices.
Then $SO(6)$ algebra generators can be constructed as
\begin{eqnarray}
  &&(\Sigma^{AB})^{\gamma}_{\alpha}=-\frac{1}{4}\big(\Sigma^{A}_{\alpha\beta}
  \Sigma^{B}{}^{\beta\gamma}
  -\Sigma^{B}_{\alpha\beta}\Sigma^{A}{}^{\beta\gamma}\big) ,
\end{eqnarray}
Defining
\begin{equation}
V_{\alpha\beta}=i\,\Sigma^A_{\alpha\beta}V_A\,,\qquad
V^{\alpha\beta}={1\over2}\epsilon^{\alpha\beta\gamma\delta}V_{\gamma\delta} ,
\end{equation}
one finds that \eqref{3.5} implies the constraint
\begin{equation}
V^{\alpha\gamma}V_{\beta\gamma}=\delta^\alpha_\beta .
\end{equation}
Using the symmetries of the l.h.s. and the fact that $\Sigma^{AB}$ is traceless,
leads to the identity
\begin{align}
  &\epsilon_{ABCDMN}\,\Sigma^{A}_{\alpha\beta}\Sigma^{B}_{\gamma\delta}
  (\Sigma^{CD})_{\lambda}^\rho(\Sigma^{MN})_{\mu}^\nu\nonumber\\
&\quad = 4i\big[\epsilon_{\alpha\beta\lambda\mu}\delta_{\gamma\delta}^{\rho\nu}
-\epsilon_{\gamma\delta\lambda\mu}\delta_{\alpha\beta}^{\rho\nu}
-\frac{1}{2}\epsilon_{\alpha\beta\gamma\lambda}\delta_{\delta}^{\nu}\delta_{\mu}^{\rho}
+\frac{1}{2}\epsilon_{\alpha\beta\delta\lambda}\delta_{\gamma}^{\nu}\delta_{\mu}^{\rho}
+\frac{1}{2}\epsilon_{\gamma\delta\alpha\lambda}\delta_{\beta}^{\nu}\delta_{\mu}^{\rho}
\label{A.13}\\
&\qquad  -\frac{1}{2}\epsilon_{\gamma\delta\beta\lambda}\delta_{\alpha}^{\nu}\delta_{\mu}^{\rho}
-\frac{1}{2}\epsilon_{\alpha\beta\gamma\mu}\delta_{\delta}^{\rho}\delta_{\lambda}^{\nu}
+\frac{1}{2}\epsilon_{\alpha\beta\delta\mu}\delta_{\gamma}^{\rho}\delta_{\lambda}^{\nu}
+\frac{1}{2}\epsilon_{\gamma\delta\alpha\mu}\delta_{\beta}^{\rho}\delta_{\lambda}^{\nu}
-\frac{1}{2}\epsilon_{\gamma\delta\beta\mu}\delta_{\alpha}^{\rho}\delta_{\lambda}^{\nu}\big] 
\nonumber
\end{align}
with
\begin{equation}
\delta^{\alpha\beta}_{\gamma\delta}=\delta^\alpha_\gamma\delta^\beta_\delta-
\delta^\alpha_\delta\delta^\beta_\gamma\,.
\end{equation}
Other useful identities are
\begin{align}
&\epsilon_{ABCDMN}(\Sigma^{AB})^\alpha_\beta(\Sigma^{CD})^\gamma_\delta
(\Sigma^{MN})^\mu_\nu~=-16 ~i~ d^{\alpha\gamma\mu}_{\beta\delta\nu} ,\\
\noalign{\vskip.2cm}
&(\Sigma^A)_{\alpha\beta}(\Sigma_A)_{\gamma\delta}=-2~\epsilon_{\alpha\beta\gamma\delta} ,
\end{align}
and
\begin{equation}
 h_{\alpha\beta}\wedge h_{\gamma\delta}
=-{1\over2}\big( V_{\alpha\gamma}\,H^{(2)}_{\beta\delta}
-V_{\beta\gamma}\,H^{(2)}_{\alpha\delta}-V_{\alpha\delta}\,H^{(2)}_{\beta\gamma}
+V_{\beta\delta}H^{(2)}_{\alpha\gamma}\big) .
\end{equation}
For an antisymmetric one-form $h_{\alpha\beta}$ with
$V^{\alpha\beta}h_{\alpha\beta}=0$ (e.g. for $h_{\alpha\beta}=D V_{\alpha\beta}$)
and a two-form $f^\alpha_\beta$
one finds the identity
\begin{eqnarray}
&&\frac{1}{2}(V^{\beta\alpha}h_{\mu\sigma}-h^{\beta\alpha}V_{\mu\sigma})f^{\mu}_{\beta}
\wedge f^{\sigma}_{\alpha}+ V^{\alpha \lambda}h_{\mu\lambda}f^{\mu}_{\sigma}
\wedge f^{\sigma}_{\alpha}=V^{\alpha \lambda}h_{\mu\lambda}f^{\mu}_{\alpha}
\wedge f^{\sigma}_{\sigma}. \quad\quad\label{A.16}
\end{eqnarray}
We will also use
\begin{align}\label{A16}
&V_{\alpha\beta}V_{\gamma\delta}+V_{\alpha\gamma}V_{\delta\beta}
+V_{\alpha\delta}V_{\beta\gamma}=\epsilon_{\alpha\beta\gamma\delta}\nonumber\\
&V_{\rho_{1}\rho_{2}}V^{\delta_{1}\delta_{2}}=
\epsilon^{\delta_{1}\delta_{2}\tau_{1}\tau_{2}}V_{\tau_{1}\rho_{1}}V_{\tau_{2}\rho_{2}}
+\delta^{\delta_{1}\delta_{2}}_{\rho_{1}\rho_{2}} .
\end{align}

\section*{Appendix B: Topological Actions and Coset Construction}
\setcounter{equation}{0}\setcounter{footnote}{0}
\renewcommand{\theequation}{B.\arabic{equation}}
\quad

In this appendix we describe an attempt to construct an action for the spin two and
spin three fields with manifest $SU(10)$ symmetry, generalizing the coset space
construction described in Section 4.
While the symmetry is manifest we will find that this construction leads
to unwanted mixed terms between the spin two and spin three fields at the quadratic level.

We begin with an alternative way to write the action \eqref{3.7}
in $SU(4)$ invariant form. Note
that the integrand in \eqref{3.7} is just the $SO(6)$ invariant trace of three elements of
the $SO(6)$ algebra or, equivalently, that
$\epsilon_{ABCDEF}$ is the $d$-symbol of $SO(6)\simeq SU(4)$.
With this observation it is immediate how to generalize the topological action for any
Lie group $G$:
\begin{equation}\label{B.16}
S_{G}\sim\int_{M_{5}}d_{\Omega\Theta\Lambda}B^{\Omega}\wedge F^{\Theta}\wedge F^{\Lambda} ~,
\end{equation}
where capital Greek indices  $\Gamma,\Theta,\Lambda\dots\in\{1,\dots,{\rm dim}(G)\}$.
The crucial point of this construction is the choice of
the coset $G/H$ whose element will be used for the
construction of the $G$ covariant one-form $B^{\Omega}$.
In the case of
$G=SO(6)$ we have $H=SO(5)$ and the compensator field is an element of the
five-sphere. Equivalently for the same system, if $G=SU(4)$
we identify the stabilizer group $H=Sp(4)\simeq SO(5)$ and the
compensator ${  V}^{\alpha\beta}$ is an element of the coset
\begin{equation}\label{B.17}
  SU(4)/Sp(4) ~,
\end{equation}
and is expressed as an antisymmetric $SU(4)$ tensor constrained by (\ref{3.10}).
Then the $SU(4)$ algebra valued one-form can be constructed as (\ref{3.15}) and
the general action (\ref{B.16}) transforms into (\ref{3.13}).
Note also that in the same fashion as we fixed the gauge using local
$SO(6)$ rotations,
\begin{eqnarray}
  &&{  V}^{A}=({  V}^{a},{  V}^{6}), \quad (a=1,\dots, 5) ,\nonumber\\
  &&{  V}^{(0)A}=(0,1) ,\label{B.18}
\end{eqnarray}
in the $SU(4)$ formulation, we can bring the
compensator field ${  V}_{\alpha\beta}(x)$ to the constant symplectic form
${  V}_{\alpha\beta}^{(0)}$, leaving an unbroken symmetry $Sp(4)$.
The relation corresponding to (\ref{B.18}) is
\begin{equation}\label{B.19}
  {  V}_{\alpha\beta}(x)={  V}_{\alpha\beta}^{(0)}=i\,\Sigma^{6}_{\alpha\beta} .
\end{equation}

We now extend the discussion to a possible compensator field for the unfied discussion
of spin $2$ and spin $3$ cases based on the $SU(10)$ algebra.
To this end we
consider an action with gauge group  $SU(10)$ with the special embedding of $SU(4)$
discussed in Section 2.
This means that we identify in (\ref{B.16}) the field strength  $F^{\Lambda}$ with the $SU(10)$
field strength (\ref{2.3}). In other words we replace the indices
$\Gamma,\Theta,\Lambda,\dots$ by two symmetrised pairs of $SU(4)$ indices
${}^{\alpha\beta}_{\gamma\delta}$ with the corresponding $SU(10)$ rule for taking the trace,
e.g. using the $d$-symbol \eqref{A.3}
\begin{equation}\label{B.20}
  S_{SU(10)}= \int_{M_{5}}B^{\alpha\beta}_{\mu\nu}\wedge
  {F}^{\mu\nu}_{\lambda\rho}\wedge{F}^{\lambda\rho}_{\alpha\beta} ,
\end{equation}
${F}^{\alpha\beta}_{\gamma\delta}$ was defined in \eqref{2.3}. It remains
to define the possible coset space and compensator, and to construct
an $SU(10)$ covariant one-form
\begin{eqnarray}\label{B.21}
&&{  B}^{\alpha\beta}_{\gamma\delta}\,,\qquad {  B}^{\alpha\beta}_{\alpha\beta}=0,\\
\noalign{\vskip.2cm}
&&\delta{  B}^{\alpha\beta}_{\gamma\delta}
 ={  B}^{\alpha\beta}_{\lambda\rho}\epsilon^{\lambda\rho}_{\gamma\delta}
-{  B}^{\lambda\rho}_{\gamma\delta}\epsilon^{\alpha\beta}_{\lambda\rho} .\nonumber
\end{eqnarray}
Searching for a suitable stabilizer for the coset $G/H$ constructed from  $G=SU(10)$,
we arrive at $H=SO(10)$. This choice of compensator allows the background value described
by the SU(4)/Sp(4) coset construction. This  property we use below in the analysis
of the linearized limit. From
\begin{eqnarray}
  G/H &=& SU(10)/SO(10) ~,\label{B.23}\\
  {\rm dim} (G/H) &=&  {\rm dim}(SU(10)) -{\rm dim}(SO(10))={54}~. \nonumber
\end{eqnarray}
we conclude that the compensator should appear as a ${54}$-dimensional
representation of $SO(10)$. For $SU(10)$ covariance of ${  B}$ or,
equivalently, for $SU(10)$ invariance of the action \eqref{B.20},
this representation should be expressed as a constrained representation of $SU(10)$.
From an $SO(10)$ point of view it is a second rank \emph{symmetric} traceless
tensor with $54$ independent real components, which we can express as an $SU(10)$
object in the following way. Consider the space of complex tensors
symmetric in a pair of lower indices and its complex conjugate
tensor with upper indices
\begin{eqnarray}
  &&{  V}_{IJ}={  V}_{JI},
  \quad \bar{{  V}}^{IJ}=\bar{{  V}}^{JI}=({  V}_{IJ})^{\ast},
  \quad I,J,\dots \in\{1,\dots 10\} .\label{B.25}
\end{eqnarray}
It has $55$ independent complex components.
The natural $SU(10)$ invariant (real) constraints
\begin{eqnarray}\label{B.26}
   &&\bar{{  V}}^{IK}{  V}_{KJ}=\delta^{I}_{J}
   \qquad \hbox{or}\quad {  V}^* {  V}={\mathbbm 1} ,\\
   &&\det({  V_{IJ}})=1 \label{B.10}
\end{eqnarray}
reduces the number of independent real components to $54$ and we can identify
this tensor with an element of the symmetric space (\ref{B.23}).
Then we can construct an $SU(10)$ covariant traceless one-form in the usual way
\begin{eqnarray}
  &&{  B}^{I}_{J}= i \bar{{  V}}^{IK}D{  V}_{KJ},\label{B.27}\\
  &&D{  V}_{KJ}=d{  V}_{KJ}-A_{(K}^{L}{  V}_{J)L} ,\nonumber
\end{eqnarray}
Moreover as opposed to the $SU(4)$ case\footnote{The identity (\ref{A.16}) relates two
possible expressions for the spin 2 action.} for $SU(10)$
we can construct one more invariant action.
Such a term can be constructed with the rank four
$d$ symbol of $SU(10)$, defined as the completely symmetrized trace of
four $SU(10)$ generators:
\begin{eqnarray}
S_{G}\sim \int_{M_{5}}d_{\Omega\Xi\Theta\Lambda}B^{\Omega\Xi}\wedge F^{\Theta}\wedge F^{\Lambda} ~.\label{B.12}
\end{eqnarray}
As before, capital Greek indices refer to the adjoint representation of $SU(10)$ and
we can replace them by an upper and a lower index refering to the fundamental representation
of $SU(10)$ and its complex conjugate, respectively, e.g.
$F^\Lambda\to F^I_J$ with
$F^I_I=0$ or by two pairs of symmetrised $SU(4)$ indices, i.e. $F^{\alpha\beta}_{\gamma\delta}$
with $F^{\alpha\beta}_{\alpha\beta}=0$.
The tensor $B$ can be realized using the $SU(10)/SO(10)$
compensator field (cf. \eqref{B.25}--\eqref{B.10})
\footnote{The traces would give the same contribution as \eqref{B.20}.}:
\begin{equation}
  B^{IK}_{JL}=\frac{i}{2}(\bar{V}^{IK}DV_{JL} - D\bar V^{IK}V_{JL})-\hbox{traces} .
\end{equation}

Replacing capital Latin  indices with symmetrized pairs of
$SU(4)$ indices as before, we arrive at the following expression for
$B^{\alpha\beta}_{\mu\nu}$ in (\ref{B.20})
\begin{eqnarray}
  {  B}^{\alpha\beta}_{\gamma\delta}&=& i\bar{{  V}}^{\alpha\beta,\lambda\rho}
  D{  V}_{\gamma\delta,\lambda\rho },\label{B.29}\\
  {  B}^{\alpha\beta}_{\alpha\beta}&=&0 ,\nonumber
\end{eqnarray}
where the $SU(10)/SO(10)$ compensator field is defined as
\begin{eqnarray}
  &&V_{\alpha\beta,\lambda\rho}={  V}_{\lambda\rho,\alpha\beta} ,\nonumber \\
  &&\bar{{  V}}^{\alpha\beta,\lambda\rho}
  = ({  V}_{\alpha\beta,\lambda\rho})^{\ast} ,\nonumber\\
  &&\bar{{  V}}^{\alpha\beta,\lambda\rho}{  V}_{\lambda\rho,\gamma\delta}
  =\delta^{\alpha\beta}_{\gamma\delta} ,\label{B.33}\\
  && \det(V_{(\alpha\beta),(\gamma\delta)})=1 .\nonumber
\end{eqnarray}
The second action (\ref{B.12}) in the $SU(4)$ covariant notation is
\begin{eqnarray}
\tilde{S}_{SU(10)} &= & \int_{M_{5}} B^{\alpha\beta,\sigma\delta}_{\mu\nu,\lambda\rho}\wedge
F^{\mu\nu}_{\alpha\beta}\wedge F^{\lambda\rho}_{\sigma\delta} ,\label{4.36}
\end{eqnarray}
where
\begin{equation}\label{B.37}
B^{\alpha\beta,\sigma\delta}_{\mu\nu,\lambda\rho}=\frac{i}{2}(\bar{V}^{\alpha\beta,\sigma\delta}
DV_{\mu\nu,\lambda\rho} - D\bar{V}^{\alpha\beta,\sigma\delta}V_{\mu\nu,\lambda\rho}) .
\end{equation}

In this case we can also use local $SU(10)$ transformations of the compensator field
and set
\begin{equation}\label{B.35}
  {  V}_{\alpha\beta,\lambda\rho}^{(0)}=\delta_{(\alpha\beta),(\lambda\rho)} .
\end{equation}
The unbroken symmetry is $SO(10)$, because the r.h.s. of (\ref{B.35})
remains invariant under $SO(10)$ rotations.

We now address the embedding of the $SU(4)/Sp(4)$
compensator ${  V}_{\alpha\beta}$ into the $SU(10)/SO(10)$ element (\ref{B.33}).
It is easy to see that the restrictions imposed by the ansatz
\begin{align}
& V_{\alpha\beta,\sigma\delta}= \frac{1}{2}(V_{\alpha\sigma}V_{\beta\delta}
+V_{\beta\sigma}V_{\alpha\delta}) ,\cr
&\bar{V}^{\alpha\beta,\sigma\delta}= \frac{1}{2}(V^{\alpha\sigma}V^{\beta\delta}
+V^{\beta\sigma}V^{\alpha\delta}) ,\label{B.36}
\end{align}
supplemented with
\begin{equation}\label{B.50}
  {  A}^{\alpha\beta}_{\mu\nu}\sim \delta^{(\alpha}_{(\mu}\omega^{\beta)}_{\nu)} ,
\end{equation}
lead to a reduction of the one-forms
\begin{align}
  {  B}^{\alpha\beta}_{\gamma\delta}=& i\bar{{  V}}^{\alpha\beta,\lambda\rho}
  D{  V}_{\lambda\rho,\gamma\delta}
  =\frac{1}{2} \delta^{(\alpha}_{(\gamma}{  B}^{\beta)}_{\delta)} ,\nonumber \label{B.38}\\
  {  B}^{\beta}_{\delta} =& i V^{\alpha\beta}D{  V}_{\alpha\delta} .
\end{align}
This means that putting the spin three gauge field to zero and using the ansatz
(\ref{B.36}), we obtain the purely gravitational action (\ref{3.13}) from the
$SU(10)$ invariant actions. This immediately shows that the equations of motion have
$AdS_{5}$ background
solutions.

Expressions (\ref{B.29}) and (\ref{B.37}) form all possible $SU(10)$ covariant one forms
which we can construct using this compensator field. Therefore the most general action
should be a linear combination
\begin{equation}\label{4.38}
  S_{SU(10)} + \kappa\, \tilde{S}_{SU(10)} ,
\end{equation}
where the relative coefficient $\kappa$ is fixed
by comparison with the free spin three action of Vasiliev \eqref{4.33}.
Trying to fix it we replace in \eqref{B.36} and \eqref{B.20} $F$ with linearized
curvatures $F^{\alpha\beta}_{1\mu\nu}=R^{\alpha\beta}_{1\mu\nu}
+\frac{1}{6}\delta^{(\alpha}_{(\mu}r^{\beta)}_{1\nu)}$ , use the $SU(4)$ restriction
\eqref{B.36} for the $SU(10)$ compensator field and replace
the covariant derivative by $D_{\omega_{0}}$. Straightforward calculation gives
\begin{eqnarray}
&&S_{SU(10)}+\kappa\, \tilde{S}_{SU(10)}\to i \int_{M^5}\frac{8}{9}(1-\kappa)
V^{\alpha \lambda}h_{\mu\lambda}
\wedge r^{\mu}_{1\delta}\wedge r^{\delta}_{1\alpha}+ S_{\rm mixed}(r_{1};R_{1})\nonumber\\
&&+i \int_{M^5}\Big[2 V^{\alpha \lambda}h_{\mu\lambda}
\wedge R^{\mu\sigma}_{1\delta_1\delta_2}\wedge R^{\delta_1\delta_2}_{1\alpha\sigma}
-2\kappa\, V^{\alpha \lambda}h_{\mu\lambda}\wedge R^{\mu\rho_1}_{1\sigma\delta_1}
\wedge R^{\sigma\rho_2}_{1\alpha\delta_{2}}V_{\rho_{1}\rho_{2}}
V^{\delta_{1}\delta_{2}}\Big] ,
\nonumber\\\label{B.51}
\end{eqnarray}
where
\begin{equation}
S_{\rm mixed}(r_{1};R_{1}) = i \int_{M^5}\Big[\frac{4}{3} V^{\alpha\sigma}h_{\mu\sigma}
\wedge r^{\beta}_{1\nu}\wedge R^{\mu\nu}_{1\alpha\beta}
+\frac{4\kappa}{3}V^{\alpha\sigma}h_{\mu\sigma}
\wedge r^{\nu}_{1\beta}\wedge R^{\mu\rho}_{1\alpha\delta}V^{\beta\delta}V_{\nu\rho}\Big] .
\label{B.52}
\end{equation}
We see that the two possible independent $SU(10)$ invariant structures
produce two independent contributions to the mixed term action (\ref{B.52}).
However there is no choice for the relative cofficient
$\kappa$ which trivializes the ``extra" field equation of motion in the second line of
(\ref{B.51}) ($\kappa=-\frac{1}{2}$, cf. \eqref{4.33})
and in the mixed term action (\ref{B.52})
($\kappa=-1$) simultaneously. This makes the correct free limit for
the coset $SU(10)$ action unreachable, at least with the ansatz (\ref{B.36}).
At the moment we do not know how to resolve this problem.

\end{document}